# The Two Loop Long Range Effect on the Proton Decay Effective Lagrangian

Takeshi Nihei and Jiro Arafune

*Institute for Cosmic Ray Research, University of Tokyo,*

*Midori-cho, Tanashi-shi, Tokyo 188 JAPAN*

**Abstract**

We calculate the two loop long range effect on the proton decay effective Lagrangian. Numerical calculation for suppression factor gives $A_L(\text{2-loop}) = 0.321$ for the value of the strong coupling constant $\alpha_s(m_Z) = 0.116$. Two loop effect to more general effective Lagrangian is also given.

## 1 Introduction

Proton decay is one of the most important prediction of grand unified theory (GUT). The main decay mode $p \to K^+ + \overline{\nu}$ in the minimal supersymmetric (SUSY) SU(5) GUT has been searched for with the underground experiments, and the lifetime bounds of this mode have been given by Kamiokande and IMB as $\tau > 1.0 \times 10^{32}$ years and $\tau > 0.62 \times 10^{32}$ years, respectively.[1] These bounds are rather close to the theoretical prediction[2][3][4] in case of the SUSY breaking parameters of less than 1 TeV and the Superkamiokande is expected to improve these bounds drastically in the near future. It is now desirable to make a more careful estimation of the theoretical bounds.

In the following we estimate the two loop QCD correction to the effective proton decay Lagrangian. This correction is estimated by solving the renormalization group equation from the weak energy scale to the low energy scale, say 1 GeV.

In order to formulate the renormalization group equation we need to estimate the two loop renormalization constant of the effective Lagrangian, which is not calculated so far though the two loop effect of the quark wavefunction, the quark masses and $\beta$-function are calculated in the literature[5]; this is why reliable calculation with the two loop approximation is not made yet. In this paper, the renormalization constant is calculated for rather general effective Lagrangian for proton decay in the two loop approximation and the renormalization group equation of the coefficient of the proton decay effective operator is given. The numerical estimation of the effect is also made.

## 2 The renormalization constant of the effective operator

The dominant proton decay effective Lagrangian at the weak scale is written as[6]

$$\mathcal{L}_{\text{eff}} = C^{ijkl}\epsilon_{\alpha\beta\gamma}(q_i^\alpha, q_j^\beta)(q_k^\gamma, \ell_l) + \text{h.c.}, \tag{1}$$

where $q$ and $\ell$ are left-handed Weyl spinors of quark and lepton field respectively. Here $\alpha, \beta, \gamma$ are color indices, $i, j, k, l$ are flavor indices and the bracket ( , ) means the inner product for Weyl spinors.

We shall also consider the proton decay Lagrangian with mixed quark chiralities, like

$$\mathcal{L}'_{\text{eff}} = D^{ijkl}\epsilon_{\alpha\beta\gamma}(\overline{q^c}_i^\alpha, \sigma^\mu q_j^\beta)(\overline{q^c}_k^\gamma, \sigma_\mu \ell_l) + \text{h.c.}, \tag{2}$$

or

$$\mathcal{L}''_{\text{eff}} = E^{ijkl}\epsilon_{\alpha\beta\gamma}(\overline{q^c}_i^\alpha, \overline{q^c}_j^\beta)(q_k^\gamma, \ell_l) + \text{h.c.}, \tag{3}$$

to discuss the two loop corrections to different four fermion vertices. Here $\sigma^\mu$ denotes the Pauli matrices and $q^c$ is the Weyl spinor for antiquark field.

The long range effect of renormalization on the coefficient $C^{ijkl}$ in Eq.(1) is to be evaluated by the ratio of its value at 1GeV to that at the weak scale $m_Z \simeq 90$GeV, that is,
$$R_L = \frac{C^{ijkl}(1\text{GeV})}{C^{ijkl}(m_Z)}. \tag{4}$$

In literatures, however, the following quantity is more frequently used:
$$A_L = \frac{\left[\dfrac{C^{ijkl}}{m_u m_d}\right]_{\mu = 1\text{GeV}}}{\left[\dfrac{C^{ijkl}}{m_u m_d}\right]_{\mu = m_Z}}, \tag{5}$$

where $\mu$ represents the renormalization point and $m_u$ or $m_d$ means the up-quark or the down-quark mass respectively.

The one loop correction has already been given as[2]
$$\begin{aligned}
A_L(1\,\text{loop}) &= \left(\frac{\alpha_s(m_b)}{\alpha_s(m_Z)}\right)^{-\frac{18}{23}} \left(\frac{\alpha_s(m_c)}{\alpha_s(m_b)}\right)^{-\frac{18}{25}} \left(\frac{\alpha_s(1\text{GeV})}{\alpha_s(m_c)}\right)^{-\frac{18}{27}} \\
&= 0.45.
\end{aligned} \tag{6}$$

Here the strong coupling constant at the renormalization point $\mu = m_Z (m_Z \simeq 90\text{GeV})$ is taken as $\alpha_s(m_Z) = 0.116$.

We take account of only QCD correction since the strong coupling constant is much larger than the weak or electromagnetic coupling constants. Up to the two loop level, there are six different types of Feynman diagrams for the QCD correction as given in the Fig.'s 1(a), 1(b), $\cdots$, 1(f).

Each type of diagram in Fig.1 is calculated by straightforward calcutations and the result is given in Table 1. In the table, $\overline{\text{MS}}$ scheme is adopted[7],
$$\frac{1}{\bar{\epsilon}} = \frac{1}{\epsilon} - \gamma + \ln(4\pi), \tag{7}$$

where $n = 4 - 2\epsilon$ ( $n$-dimensional space-time ) and $\gamma$ is an Euler number.

In Table 1, we take $A = g_s^2/(4\pi)^2 = \alpha_s/(4\pi)$ ( $g_s$ is a strong coupling constant ) and $\eta$ as the gauge parameter in the gluon propagator in the following form:
$$D_{F\alpha\beta}^{\mu\nu}(k^2) = \frac{-i\delta_{\alpha\beta}}{k^2}\left(g^{\mu\nu} - \eta\frac{k^\mu k^\nu}{k^2}\right). \tag{8}$$

From Table 1, we obtain the renormalization constant $Z_C$ as

$$Z_C = 1 - 4\frac{A}{\bar{\epsilon}} + \frac{A^2}{\bar{\epsilon}}\left(-\frac{74}{3} + \frac{8N_F}{9} + \frac{29}{9}\eta - \frac{3}{4}\eta^2 + \Delta\right)$$
$$+ \frac{A^2}{\bar{\epsilon}^2}\left(25 - \frac{2N_F}{3} - \frac{31}{2}\eta + \frac{7}{2}\eta^2\right), \quad (9)$$

including the one loop effect. Here $\Delta = 0$ for the left-handed ( or right-handed ) quarks and lepton as in Eq.(1), while $\Delta = -10/3$ for the mixed fermion chiralities as in Eq.(2), Eq.(3).

## 3 The renormalization group equation of $C^{ijkl}$

The quark wavefunction renormalization constant $Z_q$, QCD $\beta$-function $\beta$, quark mass renormalization factor $\gamma_m$ are calculated in the literature[5] as followes;

$$Z_q = 1 - \frac{A}{\bar{\epsilon}}\alpha T^2 - \frac{A^2}{\bar{\epsilon}}T^2\left\{\left(\frac{\alpha^2}{8} + \alpha + \frac{25}{8}\right)C_2 - N_F t - \frac{3}{4}T^2\right\}$$
$$+ \frac{A^2}{\bar{\epsilon}^2}T^2\left\{\left(\frac{\alpha^2}{4} + \frac{3\alpha}{4}\right)C_2 + \frac{\alpha^2}{2}T^2\right\}, \quad (10)$$

$$\beta(g_s) = -Ag_s\left(\frac{11}{3}C_2 - \frac{4tN_F}{3}\right) - A^2 g_s\left(\frac{34}{3}C_2^2 - 4N_F tT^2 - \frac{20}{3}N_F tC_2\right), \quad (11)$$

$$\gamma_m(g_s) = 8A + 8A^2\left(\frac{101}{6} - \frac{5N_F}{9}\right). \quad (12)$$

with $N_F$ being the number of quark flavors, $\alpha = 1 - \eta$, $C_2 = 3$, $T^2 = \frac{4}{3}$ and $t = \frac{1}{2}$.

The running of $g_s$ and the quark mass $m_q$ are given by

$$\mu\frac{d}{d\mu}g_s = \beta(g_s),$$
$$\mu\frac{d}{d\mu}m_q = -\gamma_m(g_s)m_q. \quad (13)$$

Combining Eq.(9) and Eq.(10), we get the renormalization group equation of the coefficient $C^{ijkl}$ as

$$\mu\frac{d}{d\mu}C^{ijkl} = -\left\{\mu\frac{d}{d\mu}\log(Z_C Z_q^{-\frac{3}{2}})\right\}C^{ijkl}$$

$$= -\left\{4A + A^2\left(\frac{14}{3} + \frac{4N_F}{9} + \Delta\right) + O(A^3)\right\} C^{ijkl}. \tag{14}$$

In the following we take $\Delta = 0$ unless otherwise stated, since the dominant decay mode in the minimal SUSY SU(5) GUT is derived from the Lagrangian in Eq.(1) with left-handed quarks and lepton.

Solving Eq.(14), one can get the two loop long range effect on the proton decay effective Lagrangian between weak scale and 1GeV:

$$\begin{aligned}
A_L(2\,\text{loop}) &= \left(\frac{\alpha_s(m_b)}{\alpha_s(m_Z)}\right)^{-\frac{18}{23}} \left(\frac{\alpha_s(m_c)}{\alpha_s(m_b)}\right)^{-\frac{18}{25}} \left(\frac{\alpha_s(1\text{GeV})}{\alpha_s(m_c)}\right)^{-\frac{18}{27}} \\
&\quad \times \left(\frac{\alpha_s(m_b) + \frac{46\pi}{58}}{\alpha_s(m_Z) + \frac{46\pi}{58}}\right)^{\frac{18}{23} - \frac{327}{116}} \times \left(\frac{\alpha_s(m_c) + \frac{50\pi}{77}}{\alpha_s(m_b) + \frac{50\pi}{77}}\right)^{\frac{18}{25} - \frac{341}{154}} \\
&\quad \times \left(\frac{\alpha_s(1\text{GeV}) + \frac{54\pi}{96}}{\alpha_s(m_c) + \frac{54\pi}{96}}\right)^{\frac{18}{27} - \frac{355}{192}}. \tag{15}
\end{aligned}$$

The result of numerical calculation is given in Table 2. Two loop correction changes $A_L$, but $R_L$ does not change very much. Uncertainty of the strong coupling constant has a great influence on these corrections. Here we give some comments on the Lagrangians with mixed chiralities. We find the one loop correction is common to all types of the local four fermion proton decay Lagrangian with three quarks and one lepton, while we find that the two loop corrections to them are common except for the diagram of the type Fig.1(e) as seen in the Table 1.

# 4  Conclusion

We have obtained the two loop correction for the long range effect on the proton decay effective Lagrangian and got the result $R_L(\text{2-loop}) = 1.33$ and $A_L(\text{2-loop}) = 0.32$ for $\alpha_s(m_Z) = 0.116$. We have found that the two loop correction of $A_L$ is appreciably large though the two loop four fermion vertex correction itself has a small contribution compared with the quark mass and gauge coupling renormalization correction. We note that the four fermion vertex correction, though found to be small, is necessary to obtain a consistent evaluation of the two loop renormalization effect. We have also found that the two loop correction to the effective proton decay Lagrangian depends on the combination of the chiralities of the four fermion fields while the one loop correction does not.

# Acknowledgement

We would like to thank T.Goto for useful discussions.

# Figure caption

**Figure 1** : The types of diagrams in the two loop level. The bold line in the center represents that we take the inner product of $q_i$ and $q_j$ and that of $q_k$ and $\ell_l$. Fig.1(a) and Fig.1(b) represent a typical two loop diagram with a quark and a gluon self-energy correction subgraph, respectively. Fig.1(c) is that with a gluon-quark vertex correction subgraph. Fig.1(d) contains a four fermion vertex correction subgraph. Fig.1(e) has gluon crossing propagators. In Fig.1(f) gluons are emitted from three quark lines and meet at the same point.

# Table captions

**Table 1** : Contribution from each type of diagrams in Fig.1 to $Z_C$. Definitions of $A(=\alpha_s/(4\pi))$, $\bar{\epsilon}$, $\eta$ ( gauge parameter ), $N_F$ and $\Delta$ are given in the text. ( $\Delta = 0$ for the left-handed ( or right-handed ) quarks and lepton as in Eq.(1), while $\Delta = -10/3$ for the mixed fermion chiralities as in Eq.(2), Eq.(3). )

**Table 2** : Numerical results for $R_L$ and $A_L$. The value of the strong coupling constant is $\alpha_s(m_Z) = 0.116$, $0.111$ and $0.121$ for (a), (b) and (c), respectively.

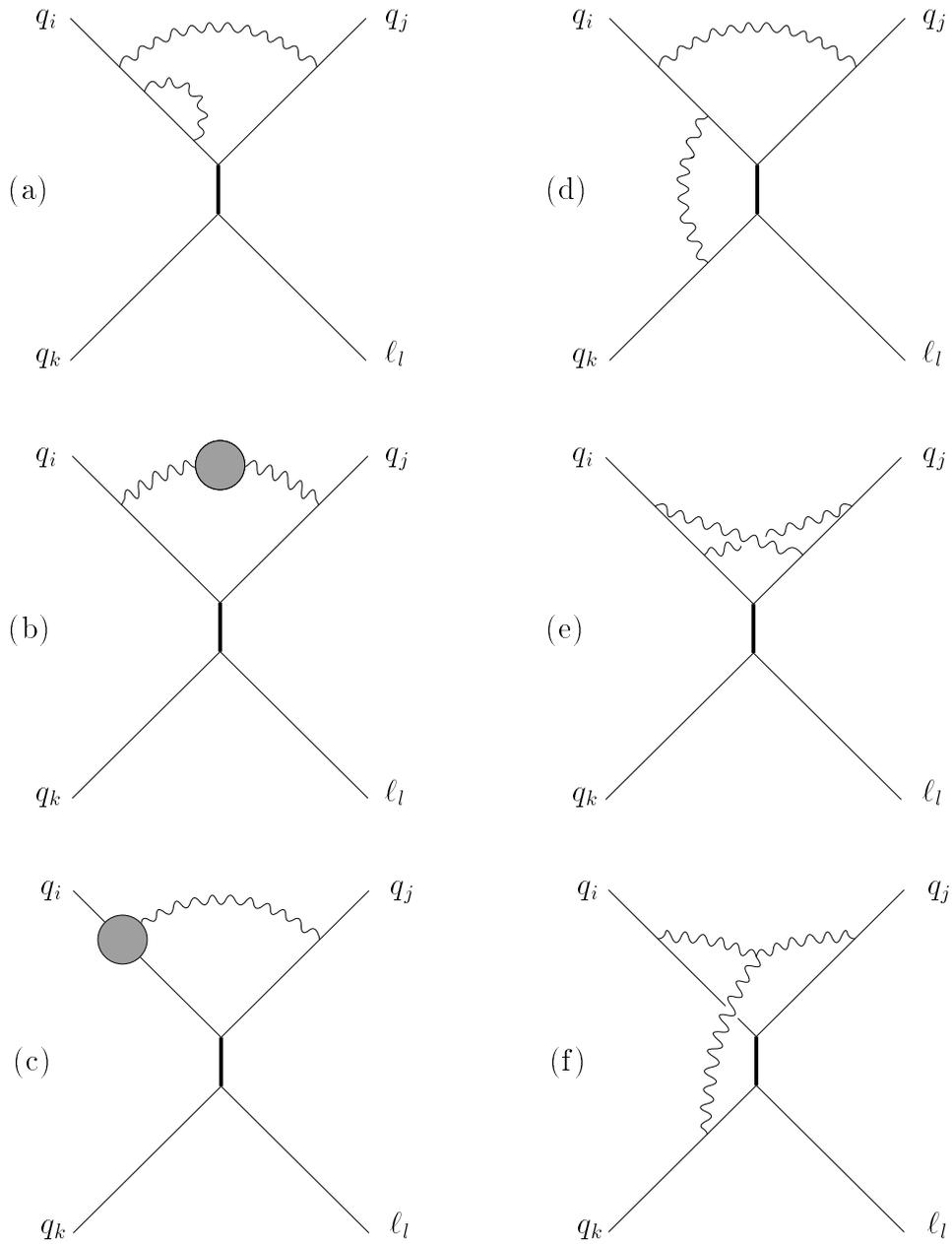

Figure 1: The types of diagram in the two loop level. The bold line in the center represents that we take the inner product of $q_i$ and $q_j$ and that of $q_k$ and $\ell_l$. Fig.1(a) and Fig.1(b) represent a typical two loop diagram with a quark and a gluon self-energy correction subgraph, respectively. Fig.1(c) is that with a gluon-quark vertex correction subgraph. Fig.1(d) contains a four fermion vertex correction subgraph. Fig.1(e) has gluon crossing propagators. In Fig.1(f) gluons are emitted from three quark lines and meet at the same point.

| type | number of diagrams | coefficient of $A^2/\bar{\epsilon}^2$ | coefficient of $A^2/\bar{\epsilon}$ |
|---|---|---|---|
| a | 6 | $-\dfrac{8}{3}(1-\eta)(2-\eta)$ | $\dfrac{16}{9}(1-\eta)$ |
| b | 3 | $5 - \dfrac{2N_F}{3} + \dfrac{3}{2}\eta$ | $-\dfrac{26}{3} + \dfrac{8N_F}{9} + \dfrac{7}{2}\eta - \dfrac{3}{4}\eta^2$ |
| c | 6 | $\dfrac{52}{3} - 17\eta + \dfrac{25}{6}\eta^2$ | $-\dfrac{112}{9} - \dfrac{2}{9}\eta$ |
| d | 9 | $2(2-\eta)^2$ | $-\dfrac{16}{3} + \dfrac{8}{3}\eta$ |
| e | 3 | 0 | $\dfrac{10}{3}\eta + \Delta$ |
| f | 1 | 0 | 0 |
| total | 28 | $25 - \dfrac{2N_F}{3} - \dfrac{31}{2}\eta + \dfrac{7}{2}\eta^2$ | $-\dfrac{74}{3} + \dfrac{8N_F}{9} + \dfrac{29}{9}\eta - \dfrac{3}{4}\eta^2 + \Delta$ |

Table 1: Contribution from each type of diagrams in Fig.1 to $Z_C$. Definitions of $A(=\alpha_s/(4\pi))$, $\bar{\epsilon}$, $\eta$ ( gauge parameter ), $N_F$ and $\Delta$ are given in the text. ( $\Delta = 0$ for the left-handed ( or right-handed ) quarks and lepton as in Eq.(1), while $\Delta = -10/3$ for the mixed fermion chiralities as in Eq.(2), Eq.(3). )

(a) $\alpha_s(m_Z) = 0.116$

|        | $R_L$ | $A_L$ |
|--------|-------|-------|
| 1-loop | 1.31  | 0.449 |
| 2-loop | 1.34  | 0.321 |

(b) $\alpha_s(m_Z) = 0.111$

|        | $R_L$ | $A_L$ |
|--------|-------|-------|
| 1-loop | 1.28  | 0.476 |
| 2-loop | 1.31  | 0.363 |

(c) $\alpha_s(m_Z) = 0.121$

|        | $R_L$ | $A_L$ |
|--------|-------|-------|
| 1-loop | 1.33  | 0.421 |
| 2-loop | 1.38  | 0.277 |

Table 2: Numerical results for $R_L$ and $A_L$. The value of the strong coupling constant is $\alpha_s(m_Z) = 0.116$, 0.111 and 0.121 for (a), (b) and (c), respectively.